\documentclass[aps,english,nofootinbib,notitlepage,showpacs,longbibliography]{revtex4-1}
\usepackage{ae,aecompl}
\usepackage[T1]{fontenc}
\usepackage[latin9]{inputenc}
\usepackage{babel}
\usepackage{units}
\usepackage{amsbsy}
\usepackage{amstext}
\usepackage{amssymb}
\usepackage{esint}
\usepackage[unicode=true]
 {hyperref}
\usepackage{breakurl}

\makeatletter

\newcommand{\binom}[2]{{#1 \choose #2}}

\makeatother

\begin{document}

\title{Hermiticity of the Volume Operators in Loop Quantum Gravity}

\author{S. Ariwahjoedi$^{1}$, I. Husin$^{1}$, I. Sebastian$^{1}$, F. P. Zen$^{1,2}$\vspace{1mm}
}

\affiliation{$^{1}$Theoretical Physics Laboratory, THEPI Division, Institut Teknologi
Bandung, Jl. Ganesha 10 Bandung 40132, West Java, Indonesia.\\
$^{2}$Indonesia Center for Theoretical and Mathematical Physics (ICTMP),
Indonesia.}

\begin{abstract}
\noindent 

The aim of this article is to provide a rigorous-but-simple
steps to prove the hermiticity of the volume operator of Rovelli-Smolin
and Ashtekar-Lewandowski using the angular momentum approach, as well
as pointing out some subleties which have not been given a lot of
attention previously. Besides of being hermitian, we also prove that
both volume operators are real, symmetric, and positive semi-definite,
with respect to the inner product defined on the Hilbert space over
SU(2). Other special properties follows from this fact, such as the
possibility to obtain real orthonormal eigenvectors. Moreover, the
matrix representation of the volume operators are degenerate, such
that the real positive eigenvalues always come in pairs for even dimension,
with an additional zero if the dimension is odd. As a consequence,
one has a freedom in choosing the orthonormal eigenvectors for each
2-dimensional eigensubspaces. Furthermore, we provide a formal procedure
to obtain the spectrum and matrix representation of the volume operators.
In order to compare our procedure with the earlier ones  existing
in the literature, we give explicit computational examples for the
case of monochromatic quantum tetrahedron, where the eigenvalues agrees
with the standard earlier procedure. 

\end{abstract}
\maketitle

\section{Introduction}

Loop Quantum Gravity (LQG) has been a fruitful field of research after
these three decades. The birth of LQG was initiated by the discovery
of the 'new variables' by Ashtekar \cite{Ashtekar1}, which reveal
a new path on the canonical quantization of gravity, and as an additional
advantage, the Hamiltonian constraint (given a special condition)
could be written in polynomial form \cite{Ashtekar2}. As a consequence
arising from the quantization, space are discrete in the Planck scale,
which is reflected by the discrete spectrum of area and volume operator
\cite{Carlo1,Carlo2}. The origin of the discreteness could be traced
from the fact that the Hilbert space of states is constructed over
the space of SU(2) connection, where the dependence on the connection
is inserted through the holonomy of SU(2), i.e., the cylindrical functions.
With this proposal in hand, one could obtain the candidate of Hilbert
space of quantum gravity equipped with the Ashtekar-Lewandowski measure,
as proven in \cite{Ashtekar3,Lewandowski,Lewandowski2}. Due to the
Peter-Weyl theorem, one could construct the basis on the Hilbert space
$\mathcal{H}_{\Gamma}=L_{2}\left[\textrm{SU(2)},\textrm{d}\mu_{\textrm{Haar}}\right]$
of a graph $\Gamma$, defined by a collection of intersecting loops,
from the irreducible representation of SU(2) in $(2j+1)$-dimension
\cite{Carlo5,Kristina}. For the kinematical regime, where the Gauss
constraint is taken into account, one has a spin-network state: a
gauge invariant state of a graph labeled by the spin representation
of SU(2).

There exists two versions of volume operators which differ by the
regularization procedure. The first is due to Rovelli and Smolin \cite{Carlo1}
and the later is due to Ashtekar and Lewandowski \cite{Ashtekar4,Ashtekar5}.
These operators are constructed from a triple surface integral over
the triads, which, using the terms in \cite{Eugenio}, is called as
the \textit{fluxization} of the volume. This cause the dimension of
the operators to be in the order of $L^{6}$, and to correctly describe
the volume of a 3-dimensional region of space, one needs to take their square root. Both of the operator
are well-defined in the sense that they converge to the classical
volume formula in the continuum limit \cite{Ashtekar5,Eugenio}, and
analysis on their properties could be found in considerable amount
on the literatures. However, there exists several subleties which
become a concern in this article, as some of them are already pointed
out in \cite{kristina1,Kristina2}. Another new volume operator had
been introduced by Yang and Ma \cite{Ma1}, where the regularized
volume is constructed from a triple line integral over cotriads, instead
of surfaces integral as their previous predecessors. This cause the
dimension of the new alternative volume operator to be in the order
of $L^{3}$, hence one does not need to take the square root to obtain
the correct operator of volume. The consistency and spectrum of the new alternative operator has not been studied in great detail, since the operator was introduced so recently. Its properties are an interesting subject to pursue, but
remains in the outer scope of this article.

As shown in \cite{Flori}, the expectation value of
Ashtekar-Lewandowski (and hence the Rovelli-Smolin) volume operator
coincides with the classical volume for the coherent states only for
6-valent vertices, i.e., the graph defines only cubical topology.
For $n\neq6$, the quantum states do not admit a correct semi-classical
limit, given specific coherent states in \cite{Winkler} and \cite{Coherent}.
To solve the criticism, a more general \textit{polyhedral}
volume operator was introduced by Bianchi, Dona', and Speziale in
\cite{Simone}. The polyhedral volume operator is constructed following
the procedure in \cite{Glauber,Sudarshan}, such that in the semi-classical
limit, it gives rise to any $n$-faces polyhedra, which is independent
from the definition of the coherent states (the construction in \cite{Simone}
particularly, use the Livine-Speziale coherent states \cite{Laurent}).
However, the complete spectrum of this operator has not been studied
in great detail. Due to this reason, in this article, we will only
consider the Rovelli-Smolin and Ashtekar-Lewandowski volume operators.

The importance of the volume operator in LQG varies from the needs
of studying the properties of quanta of space, to other related operators
such as the length operator \cite{Eugenio,Ma2}, Hamiltonian constraint
operator \cite{Thiemann,Carlo3,Ma3}, and Master constraint
operator \cite{Master}. It is also used in the procedure to couple
matters with gravitational field \cite{Hossain,Alesci}. In the LQG
literatures, it was already mentioned that the volume operator is
hermitian \cite{Eugenio,Carlo4,Thimanclosed,Brunemann1}, even some
articles had proven its hermiticity \cite{Carlo4,Thimanclosed,Brunemann1}.
However, the proof was not done in a formal and rigorous mathematical
manner, which in our opinion is not satisfying and transparent enough
to give a clear first sight, particularly for researchers and students
new to LQG. Therefore, the aim of this article is to provide a rigorous-but-simple
steps to prove the hermiticity of the volume operator of Rovelli-Smolin
and Ashtekar-Lewandowski, as well as pointing out some subleties which
has not been given a lot of attention previously. Interestingly, we
could also show that both the volume operators are real, symmetric,
and positive semi-definite, with respect to the inner product defined
on the Hilbert space over SU(2). In fact, the symmetricity of the
Rovelli-Smolin volume operator, based on the graphical method of the
spin-network \cite{Carlo5,Carlo4,JinsongYang}, had been proven in
\cite{Carlo4}. Our work in this article could be interpreted as a
complement to the result obtained in \cite{Carlo4}, in the sense
that the symmetricity of the volume operators is proven from the angular
momentum approach. The equivalency between these two approaches is
supported by the result.

The article is organized as follows. In Section II, we briefly review
the two well-known volume operators in LQG; these operators are rewritten
using a common notation in order to discover easily their similarity
and differences. Section III consists the proof of the hermiticity
of the volume operators; this section begins with the introduction
of several important mathematical definitions, followed by some claims
useful for the proof. We also highlight several subleties concerning
the operators. In addition to the hermiticity condition, we show that
the volume operators are real, symmetric, and positive semi-definite,
with respect to the inner product defined on $\mathcal{H}_{\Gamma}=L_{2}\left[\textrm{SU(2)},\textrm{d}\mu_{\textrm{Haar}}\right]$.
Other special properties follows from this fact, such as the possibility
to obtain real orthonormal eigenvectors, the occurence of degenerates
eigenvalues, and the freedom to choose the eigenvectors for the eigensubspaces
due to the degeneracies in the spectrum of volume operators. In Section
IV, we provide explicit calculations of the matrix representation
of volume operators and their spectrum for a quantum tetrahedron,
both for the ground state monochromatic (GSM) and first excited monochromatic
(FEM) case. Finally we give a conclusion and remarks on this subject.

\section{The Volume Operator}

In this section, we will review the well-known results on the volume
operators in LQG. There are two types of volume operator, the one
derived by Rovelli-Smolin, which are labeled by $\hat{V}_{RS}$ \cite{Carlo1,Eugenio},
and the one by Ashtekar-Lewandowski, labeled as $\hat{V}_{AL}$ \cite{Ashtekar4,Ashtekar5}.
The main differences between these two operators are the following:
(1) the constants in front of the operators, (2) the way the operator
sums up the variables for each link, and (3) the absence of the sign
factor in one of the operator. These differences are a consequence
of different regularization schemes applied to the volume operator,
but however they start from the same classical definition of volume
as follows: 
\begin{equation}
V_{\mathcal{R}}(x)=\intop_{\mathcal{R}}\textrm{vol}=\intop_{\mathcal{R}}d^{3}x\sqrt{q(x)},\label{eq: volume klasik}
\end{equation}
with $q(x)$ is the determinant of metric $\boldsymbol{q}$ on 3D
foliation $\Sigma$, and $\mathcal{R}$ is a 3D region on $\Sigma$.
Writing $\boldsymbol{q}$ in terms of densitized triads $E$, (\ref{eq: volume klasik})
could be directly quantized by promoting the triads to an operator
$E_{i}^{a}\rightarrow\hat{E}_{i}^{a}$, giving the (continuous) volume
operator in the connection representation:
\begin{equation}
\hat{V}_{\mathcal{R}}(x)=\intop_{\mathcal{R}}d^{3}x\sqrt{\left|\frac{1}{3!}\varepsilon^{ijk}\varepsilon_{abc}\hat{E}_{i}^{a}(x)\hat{E}_{j}^{b}(x)\hat{E}_{k}^{c}(x)\right|}.\label{eq:2.6 operator volume}
\end{equation}
A regularization is needed to obtain a volume operator free from singularities
\cite{Carlo5,Kristina}.

\subsection{Rovelli-Smolin Volume Operator}

The regularized Rovelli-Smolin volume is constructed by considering
a quantity such that in the continuum limit it converges to the classical
version of (\ref{eq:2.6 operator volume}). The derivation here is
based on \cite{Eugenio}. A region $\mathcal{R}$ is approximated
by a set of cubic-cell $\mathcal{R}_{\alpha}$ with the edge length
$\triangle x$, such that the region $\mathcal{R}\subset\cup_{\alpha}\mathcal{R}_{\alpha}$.
The volume of region $\mathcal{R}$ can be approximated by the Riemann
sum of the cubic-cells volumes, namely, $\sum_{\alpha}\textrm{vol}\left(\mathcal{R}_{\alpha}\right)$.
One considers the following triple surface integral of the cubes as
follows:
\begin{eqnarray}
W_{\triangle x}\left(x_{\alpha}\right) & = & \frac{1}{8\times3!}\frac{1}{\left(\Delta x\right)^{6}}\intop_{\partial\mathcal{R}_{\alpha}}d^{2}\boldsymbol{\sigma}\intop_{\partial\mathcal{R}_{\alpha}}d^{2}\boldsymbol{\sigma}'\intop_{\partial\mathcal{R}_{\alpha}}d^{2}\boldsymbol{\sigma}''\:\times..\nonumber \\
 &  & ..\times\left|T_{x_{I}}^{ijk}\left(\boldsymbol{\sigma},\boldsymbol{\sigma}',\boldsymbol{\sigma}''\right)E_{i}^{a}\left(\boldsymbol{\sigma}\right)n_{a}\left(\boldsymbol{\sigma}\right)E_{j}^{b}\left(\boldsymbol{\sigma}'\right)n_{b}\left(\boldsymbol{\sigma}'\right)E_{k}^{c}\left(\boldsymbol{\sigma}''\right)n_{c}\left(\boldsymbol{\sigma}''\right)\right|.\label{eq: fluks}
\end{eqnarray}
$\boldsymbol{\sigma}$ is the local coordinate on $\Sigma$, and $\boldsymbol{n}$
is the normal to surface $\partial\mathcal{R}_{\alpha},$ written
in coordinate $x^{a}=X^{a}\left(\boldsymbol{\sigma}\right)$ as:
\[
n_{a}\left(\boldsymbol{\sigma}\right)=\varepsilon_{abc}\frac{\partial X^{b}}{\partial\sigma^{1}}\frac{\partial X^{b}}{\partial\sigma^{2}}.
\]
$T_{x_{\alpha}}^{ijk}\left(\boldsymbol{\sigma},\boldsymbol{\sigma}',\boldsymbol{\sigma}''\right)$
is a function that guarantees the three fluxes to satisfies SU(2)
gauge invariance, namely:
\begin{equation}
T_{x_{\alpha}}^{ijk}\left(\boldsymbol{\sigma},\boldsymbol{\sigma}',\boldsymbol{\sigma}''\right)=\varepsilon^{i'j'k'}D^{1}\left(h_{\gamma_{x_{\alpha}\sigma}^{1}}\left[A\right]\right){}_{i'}^{\;\; i}D^{1}\left(h_{\gamma_{x_{\alpha}\sigma'}^{2}}\left[A\right]\right){}_{j'}^{\;\; j}D^{1}\left(h_{\gamma_{x_{\alpha}\sigma''}^{3}}\left[A\right]\right){}_{k'}^{\;\; k}.\label{eq:intertwin}
\end{equation}
$D^{1}\left(h_{\gamma}\left[A\right]\right)$ is the adjoint representation
of SU(2) holonomy along the loop $\gamma=\gamma_{x_{\alpha}\sigma}$,
starting at $x_{\alpha}$ in $\mathcal{R}_{\alpha}$ and ends at the
boundary $\partial\mathcal{R}_{\alpha}$. Taking the limit $\Delta x\rightarrow0$,
(\ref{eq: fluks}) becomes:
\begin{equation}
\lim_{\Delta x\rightarrow0}W_{\triangle x}\left(x_{\alpha}\right)=\frac{1}{3!}\left|\varepsilon^{ijk}\varepsilon_{abc}E_{i}^{a}\left(x_{\alpha}\right)E_{j}^{b}\left(x_{\alpha}\right)E_{k}^{c}\left(x_{\alpha}\right)\right|.\label{eq: Shrinking}
\end{equation}
Therefore, one has:
\begin{equation}
V_{\mathcal{R}}=\lim_{\Delta x\rightarrow0}\sum_{\alpha}\left(\Delta x\right)^{3}\sqrt{W_{\triangle x}\left(x_{\alpha}\right)},\label{eq: Volume Awal}
\end{equation}
and the regularized Rovelli-Smolin volume is:
\[
V_{RS}=\sum_{\alpha}\left(\Delta x\right)^{3}\sqrt{W_{\triangle x}\left(x_{\alpha}\right)}.
\]
The next step is to define a partition of surface $\partial\mathcal{R}_{\alpha}$
into square plaquette $S_{\alpha}^{I}$ such that $\partial\mathcal{R}_{\alpha}=\bigcup_{I}S_{\alpha}^{I}$.
Using this partition, one can write $W_{\triangle x}\left(x_{\alpha}\right)$
as a Riemann summation of three fluxes \cite{Eugenio,Carlo4}: 
\begin{eqnarray}
W_{\triangle x}\left(x_{\alpha}\right) & = & \frac{1}{8\times3!}\frac{1}{\left(\Delta x\right)^{6}}\sum_{I,J,K}\left|T_{x_{\alpha}}^{ijk}E_{i}\left(S_{\alpha}^{I}\right)E_{j}\left(S_{\alpha}^{J}\right)E_{k}\left(S_{\alpha}^{K}\right)\right|.\label{eq: fluks-1}
\end{eqnarray}
Promoting the fluxes into hermitian operators: $E_{i}\left(S^{I}\right)\rightarrow\hat{E}_{i}\left(S^{I}\right)=\mathbf{i}X_{i}^{I}$,
defined as the (self-adjoint) right invariant vector field on SU(2),
one immediately obtains the following relation:

\begin{equation}
\left(\Delta x\right)^{3}\sqrt{\hat{W}_{\triangle x}\left(x_{\alpha}\right)}=\sqrt{\frac{1}{8\times3!}\sum_{I,J,K}\left|\mathbf{i}\; T_{x_{\alpha}}^{ijk}X_{i,\alpha}^{I}X_{j,\alpha}^{J}X_{k,\alpha}^{K}\right|}.\label{eq:flux}
\end{equation}
$X_{I}^{i}$ is antihermitian, satisfying $\left[X_{I}^{i},X_{I}^{j}\right]=-2\epsilon^{ijk}X_{I}^{k}$.
They are related to self-adjoint Pauli matrices by $\tau^{i}:=-\frac{\mathbf{i}}{2}X^{i}$,
satisfying $\left[\tau^{i},\tau^{j}\right]=\mathbf{i}\epsilon^{ijk}\tau^{k}$.
Using the representation of angular momentum in $(2j+1)$-dimension
as follows: $\tau_{i}\mapsto\rho_{j}\left(\tau_{i}\right)=J_{i}$,
the argument inside the absolute value of (\ref{eq:flux}) could be
written as:
\begin{equation}
\mathbf{i}T_{x}^{ijk}X_{i}^{I}X_{j}^{J}X_{k}^{K}=8\;\varepsilon^{ijk}J_{i}^{I}J_{j}^{J}J_{k}^{K},\label{eq:triv}
\end{equation}
with the adjoint representation of the holonomy on (\ref{eq:intertwin})
is gauge-fixed to be trivial (this is possible, providing the geometrical
picture of the quanta of space which is flat in the interior). The
term in the RHS of (\ref{eq:triv}), following \cite{Brunemann1},
is written as:
\begin{equation}
\hat{q}_{IJK}:=\frac{4}{\mathbf{i}}\varepsilon_{ijk}J_{I}^{i}J_{J}^{j}J_{K}^{k}=\frac{4}{\mathbf{i}}J_{I}\cdot\left(J_{J}\times J_{K}\right).\label{eq:hey}
\end{equation}
$\hat{q}_{IJK}$ is the \textit{ three-hand} operator \cite{Eugenio,Carlo4}.
Therefore, the Rovelli-Smolin volume operator is:
\begin{eqnarray}
\hat{V}_{RS} & = & \sum_{\alpha}\left(\hat{v}_{RS}\right)_{\alpha}\nonumber \\
\left(\hat{v}_{RS}^{\left(1\right)}\right)_{\alpha} & = & \sqrt{\frac{1}{4}\sum_{I<J<K}\left|\mathbf{i}\left(\hat{q}_{\alpha}\right)_{IJK}\right|},\label{eq:rs1}\\
\left(\hat{q}_{\alpha}\right)_{IJK} & = & \left[\left(J_{IJ}\right)^{2},\left(J_{JK}\right)^{2}\right]_{\alpha},\label{eq:Vrs}
\end{eqnarray}
The last equality comes from relation (\ref{eq:hey}), together with
the fact that the summation in (\ref{eq:rs1}) is only over \textit{distinct}
indices $I,J,K$. This will be clear later in Section III.

Another version of Rovelli-Smolin volume operator exists, according
to \cite{kristina1,Kristina2}, where the difference is on the location
of the summation, which is inside the square root:
\begin{eqnarray}
\hat{V}_{RS} & = & \intop_{\mathcal{R}}d^{3}\boldsymbol{p}\:\hat{V}\left(\boldsymbol{p}\right)_{\gamma},\label{eq: V_RS-1-1}\\
\hat{V}\left(\boldsymbol{p}\right)_{\gamma} & = & \sum_{v\in V\left(\gamma\right)}\delta^{\left(3\right)}\left(\boldsymbol{v},\boldsymbol{p}\right)\left(\hat{v}_{RS}\right)_{v,\gamma}.\\
\left(\hat{v}_{RS}^{\left(2\right)}\right)_{v,\gamma} & = & \overline{\sum_{I,J,K}}\sqrt{\frac{C_{\textrm{reg}}}{8}\left|\mathbf{i}\varepsilon_{ijk}X_{I}^{i}X_{J}^{j}X_{K}^{k}.\right|}.\label{eq:huk}
\end{eqnarray}
Writing (\ref{eq:huk}) in terms of the angular momentum representation
and using (\ref{eq:hey}), one could write the second version of Rovelli-Smolin
operator using a common notation with the previous one:
\begin{eqnarray}
\left(\hat{v}_{RS}^{\left(2\right)}\right)_{v,\gamma} & = & \overline{\sum_{I,J,K}}\sqrt{\frac{C_{\textrm{reg}}}{4}\left|\mathbf{i}\left(\hat{q}_{v}\right)_{IJK}\right|},\label{eq:rs2}
\end{eqnarray}
with $C_{\textrm{reg}}$ is a constant to be specified by the regularization
procedure. Notice that the differences with (\ref{eq:Vrs}) is in
the location of the summation and the constant.

\subsection{The Ashtekar-Lewandowski Volume Operator}

On the other hand, the Ashtekar-Lewandowski volume operator $\hat{V}_{AL}$
is obtained from different regularization scheme. The derivation here
is based on \cite{Thimanclosed}. Let us consider a cubic-cell on
$\Sigma$ centered at point $p$, with edge length $2\Delta_{i}$.
Let the unit vectors $\hat{\boldsymbol{n}}_{i}$ be the normals to
the faces-$i$ of the cube, then the volume of the cube is: 
\[
\textrm{vol}\left(\triangle\right)=2^{3}\triangle_{1}\triangle_{2}\triangle_{3}\det\left(\hat{\boldsymbol{n}}_{1},\hat{\boldsymbol{n}}_{2},\hat{\boldsymbol{n}}_{3}\right).
\]
The coordinate of the cube is defined by $\boldsymbol{x},$ such that
the characteristic function of the cube is written as follows: 
\begin{equation}
\lambda_{\triangle}\left(\boldsymbol{x},\boldsymbol{p}\right)=\prod_{i=1}^{3}\Theta\left(\triangle_{i}-\left|\left\langle \hat{\boldsymbol{n}}_{i},\boldsymbol{x}-\boldsymbol{p}\right\rangle \right|\right).\label{eq:charai}
\end{equation}
$\Theta\left(z\right)$ is the Heaviside step function, with the condition
$\Theta\left(z\right)=0$ if $z<0$, $\Theta\left(z\right)=\frac{1}{2}$
if $z=0$, and $\Theta\left(z\right)=1$ if $z>0$. Taking the limit
$\triangle\rightarrow0$ by setting $\triangle_{i}\rightarrow0$ ,
(\ref{eq:charai}) becomes: 
\[
\lim_{\triangle\rightarrow0}\frac{1}{\textrm{vol}\left(\triangle\right)}\lambda_{\triangle}\left(\boldsymbol{x},\boldsymbol{p}\right)=\delta^{\left(3\right)}\left(\boldsymbol{x},\boldsymbol{p}\right),
\]
with $\delta^{\left(3\right)}$ is the three dimensional Dirac-delta
function. One considers the triple volume integral as follows: 
\begin{eqnarray}
E\left(\boldsymbol{p},\triangle,\triangle',\triangle''\right) & = & \frac{1}{3!}\frac{1}{\textrm{vol}\left(\triangle\right)\textrm{vol}\left(\triangle'\right)\textrm{vol}\left(\triangle''\right)}\int_{R}d^{3}\boldsymbol{x}\int_{R}d^{3}\boldsymbol{x}'\int_{R}d^{3}\boldsymbol{x}''\:\times..\nonumber \\
 &  & ..\times\;\lambda_{\triangle}\left(\boldsymbol{x},\boldsymbol{p}\right)\lambda_{\triangle'}\left(\frac{\boldsymbol{x}+\boldsymbol{x}'}{2},\boldsymbol{p}\right)\lambda_{\triangle''}\left(\frac{\boldsymbol{x}+\boldsymbol{x}'+\boldsymbol{x}''}{2},\boldsymbol{p}\right)\varepsilon^{ijk}\varepsilon_{abc}E_{i}^{a}(\boldsymbol{x})E_{j}^{b}(\boldsymbol{x}')E_{k}^{c}(\boldsymbol{x}''),\label{eq: fluks-2}
\end{eqnarray}
such that the volume of region $\mathcal{R}$ could be obtained by
taking the limit of $\triangle\rightarrow0$: 
\begin{equation}
V_{\mathcal{R}}=\lim_{\triangle\rightarrow0}\lim_{\triangle'\rightarrow0}\lim_{\triangle''\rightarrow0}\intop_{\mathcal{R}}d^{3}\boldsymbol{p}\:\sqrt{E\left(\boldsymbol{p},\triangle,\triangle',\triangle''\right)}.\label{eq: Volume Awal-1}
\end{equation}
Hence, the Ashtekar-Lewandowski regularization of volume is defined
as: 
\[
V_{AL}=\intop_{\mathcal{R}}d^{3}\boldsymbol{p}\:\sqrt{\hat{E}\left(\boldsymbol{p},\triangle,\triangle',\triangle''\right)}.
\]
Promoting the fluxes as operators, acting on the holonomy, and using
the proposal completely explained in \cite{Thimanclosed}, one has
the Ashtekar-Lewandowski volume operator as follows:

\begin{eqnarray}
\hat{V}_{AL} & = & \intop_{\mathcal{R}}d^{3}\boldsymbol{p}\:\hat{V}\left(\boldsymbol{p}\right)_{\gamma},\label{eq: V_RS-1}\\
\hat{V}\left(\boldsymbol{p}\right)_{\gamma} & = & \sum_{v\in V\left(\gamma\right)}\delta^{\left(3\right)}\left(\boldsymbol{v},\boldsymbol{p}\right)\left(\hat{v}_{AL}\right)_{v,\gamma}.\\
\left(\hat{v}_{AL}\right)_{v,\gamma} & = & \sqrt{\frac{1}{8\times3!}\left|\mathbf{i}\sum_{\binom{e_{I},e_{J},e_{K}\in E\left(\gamma\right)}{e_{I}\cap e_{J}\cap e_{K}=v}}\epsilon\left(e_{I},e_{J},e_{K}\right)\varepsilon_{ijk}X_{I}^{i}X_{J}^{j}X_{K}^{k}\right|}.\label{eq:huf}
\end{eqnarray}
Rewriting the right invariant vector fields in terms of angular momentum
representation as in the previous subsections, (\ref{eq:huf}) becomes:
\begin{eqnarray}
\left(\hat{v}_{AL}\right)_{v,\gamma} & = & \sqrt{\frac{1}{4}\left|\sum_{I<J<K}\epsilon\left(e_{I},e_{J},e_{K}\right)\mathbf{i}\left(\hat{q}_{v}\right)_{IJK}\right|}.\label{eq:AL}
\end{eqnarray}
with $\left(\hat{q}_{v}\right)_{IJK}$ is the three-hand operator
satisfying (\ref{eq:Vrs}). Notice that one of the main difference
between (\ref{eq:AL}) and (\ref{eq:rs1})-(\ref{eq:rs2}) is the
existence of the sign factor $\epsilon(e_{I},e_{J},e_{K})$, where
its value is determined by the cross product sign of the tangents
vectors of the link of the spin-network.

\section{The Hermiticity of the Volume Operator}

\subsection{Basic Definitions and Facts of Hermitian Matrix}

The confusion we encounter regarding the volume operators is on the
definition of the absolute value of an operator, in particular, a
matrix. It is not clear in the LQG literatures (at least, we could
not find any article which clearly discussed this matter) how the
absolute value acts on the arguments in (\ref{eq:rs1}), (\ref{eq:rs2}),
and (\ref{eq:AL}). However, in the mathematical literatures, such
definition exists on the subject of bounded operators \cite{Math1,Math2}.
We apply a similar definition for matrices as follows:

\textbf{Definition 1.} \textit{Given an arbitrary (not necessarily
regular) complex matrix $M,$ the absolute value of $M,$ namely $\left|M\right|$,
is defined as follows: 
\begin{equation}
\left|M\right|:=+\sqrt{M^{\dagger}M},\label{eq:abs}
\end{equation}
with $M^{\dagger}$ is the complex-conjugate of $M$.}

To understand the properties of $\left|M\right|$, one considers a
hermitian matrix, which is special case of complex matrices satisfying
$H^{\dagger}=H$. One could show that a hermitian matrix is unitarily
diagonalizable. It is a well-known fact that all the eigenvalues of
hermitian matrices are real, and vice-versa: the reality of the eigenvalues
guarantees a matrix to be hermitian. It needs to be kept in mind that
the hermiticity of an operator (or a matrix) is defined with respect
to a choice of inner product. A hermitian matrix with respect to inner
product $\left\langle ..,..\right\rangle _{1}$, in general will not
be hermitian with respect to a distinct inner product $\left\langle ..,..\right\rangle _{2}$.
Another definition concerning a hermitian matrix which is useful for
the discussion is the following:

\textbf{Definition 2}. \textit{A hermitian matrix $H$ is positive
semi-definite (or non-negative) if: 
\begin{eqnarray*}
\mathbf{x}^{\dagger}H\mathbf{x} & \geq & 0,\qquad\forall\;\mathbf{x}\in\mathbb{C}^{n},\;\mathbf{x}\neq0.
\end{eqnarray*}
and negative semi-definite (or non-positive) if: 
\[
\mathbf{x}^{\dagger}H\mathbf{x}\leq0,\qquad\forall\;\mathbf{x}\in\mathbb{C}^{n},\;\mathbf{x}\neq0.
\]
}

Notice that it is not relevant to define the notion of non-negativity
(or non-positivity) for an arbitrary complex matrix $M$ since the
quantity $\mathbf{x}^{\dagger}M\mathbf{x}$ in general is not restricted
to real numbers. As a consequence to Definition 1 and 2, one could
prove these following facts:

\textbf{Claim 1.} \textit{$M^{\dagger}M$ is hermitian and positive
semi-definite. }\textbf{Proof}: (1) $\left(M^{\dagger}M\right)^{\dagger}=M^{\dagger}M^{\dagger\dagger}=M^{\dagger}M$,
thus $M^{\dagger}M$ is hermitian. (2) $\mathbf{x}^{\dagger}M^{\dagger}M\mathbf{x}=\left(M\mathbf{x}\right)^{\dagger}M\mathbf{x}\geq0$,
with respect to a Riemannian inner product in complex space. Therefore
$M^{\dagger}M$ is positive semi-definite.

\textbf{Claim 2A.} \textit{A hermitian and positive semi-definite
matrix $T$ has non-negative eigenvalues. }\textbf{Proof}: Consider
the eigenvalue problem $T\mathbf{n}=\lambda_{n}\mathbf{n}$, with
$\left(\lambda_{n},\mathbf{n}\right)$ is the spectrum of $T,$ where
the eigenvalues $\lambda_{n}\in\mathbb{R}$. Acting on the left by
$\mathbf{n}^{\dagger}$, one has $\mathbf{n}^{\dagger}T\mathbf{n}=\lambda_{n}\mathbf{n}^{\dagger}\mathbf{n}$.
Both $\mathbf{n}^{\dagger}T\mathbf{n}\geq0$ and $\mathbf{n}^{\dagger}\mathbf{n}\geq0$,
therefore $\lambda_{n}\geq0$. 

\textbf{Claim 2B.} \textit{If all the eigenvalues of a diagonalizable
matrix $T$ are non-negative, then $T$ is a hermitian, positive semi-definite
matrix}. \textbf{Proof}: (1) Since the eigenvalues are real, $T$
is clearly hermitian which we are not going to prove. (2) Since a
hermitian matrix is unitarily diagonalizable, one has $\Lambda^{-1}T\Lambda=\lambda_{n}P_{n}$,
with $\left(P_{n}\right)_{ij}=\delta_{ni}\delta_{j}^{i}$ is the spectral
projector of $T$. Acting a non-zero vector $\mathbf{x}\in\mathbb{C}^{n}$
on the left and the right of $T$ one has:
\[
\mathbf{x}^{\dagger}T\mathbf{x}=\mathbf{x}^{\dagger}\Lambda\lambda_{n}P_{n}\Lambda^{\dagger}\mathbf{x},
\]
using the fact that $\Lambda^{-1}=\Lambda^{\dagger}.$ With $\bar{\mathbf{x}}=\Lambda^{\dagger}\mathbf{x}$,
one has $\mathbf{x}^{\dagger}T\mathbf{x}=\bar{\mathbf{x}}^{\dagger}\lambda_{n}P_{n}\bar{\mathbf{x}}.$
Moreover, 
\[
\mathbf{x}^{\dagger}T\mathbf{x}=\sum_{n}\overline{x}_{n}^{\dagger}\lambda_{n}P_{n}\bar{x}_{n}=\sum_{n}\lambda_{n}\left|\bar{x}_{n}\right|^{2}.
\]
If $\lambda_{n}\geq0$, then $T$ is positive semi-definite, since
$\bar{\mathbf{x}}$ is not a zero vector.

\textbf{Claim 3.} \textit{A hermitian and positive semi-definite matrix
$T$ admits a unique hermitian and positive semi-definite square root}.
\textbf{Proof}: Consider the diagonalization $\Lambda^{-1}T\Lambda=\lambda_{n}P_{n}$.
One could write the diagonalization as $\Lambda^{-1}\sqrt{T}\sqrt{T}\Lambda=\sqrt{\lambda_{n}}\sqrt{\lambda_{n}}P_{n}$,
or moreover:
\[
\Lambda^{-1}\sqrt{T}\Lambda\Lambda^{-1}\sqrt{T}\Lambda=\left(\Lambda^{-1}\sqrt{T}\Lambda\right)^{2}=\left(\sqrt{\lambda_{n}}P_{n}\right)^{2},
\]
using the fact that $P_{n}^{2}=P_{n}$. Finally one has $\Lambda^{-1}\sqrt{T}\Lambda=\pm\sqrt{\lambda_{n}}P_{n}$.
The following is the positive solution to the square root of $T,$
which is unique:
\[
\sqrt{T}=\Lambda\sqrt{\lambda_{n}}P_{n}\Lambda^{-1}.
\]
Since $\lambda_{n}\in\mathbb{R}$ and $\lambda_{n}\geq0$, then from
Claim 2B, $\sqrt{T}$ is hermitian and positive semi-definite.

\textbf{Claim 4.} \textit{The absolute value of an arbitrary matrix
defined as in (\ref{eq:abs}), namely $\left|M\right|,$ is hermitian
and positive semi-definite. }\textbf{Proof}: $M^{\dagger}M$ is hermitian,
and therefore diagonalizable. It is also positive semi-definite, and
hence from Claim 3, admits a unique, hermitian, positive semi-definite
square root, namely $+\sqrt{M^{\dagger}M}$ which by definition (\ref{eq:abs})
is $\left|M\right|$.

As direct consequence of Claim 3 and 4, we have as follows, the main
important fact which will be used to prove the hermiticity of the
volume operators:

\textbf{Claim 5.} \textit{The absolute value of an arbitrary matrix,
namely $\left|M\right|,$ admits a unique hermitian and positive semi-definite
square root, namely $\sqrt{\left|M\right|}$. }The proof for this
is clear.

\subsection{Subleties on the Volume Operators}

Let us return to the volume operators, which are listed as follows
(the indices $\alpha,$ and $v,\gamma$ are neglected for simplicity,
and the constants are written in $Z$'s):
\begin{eqnarray}
\hat{v}_{RS}^{\left(1\right)} & = & \sqrt{Z_{1}\sum_{I,J,K}\left|\mathbf{i}\hat{q}_{IJK}\right|},\label{eq:k}\\
\hat{v}_{RS}^{\left(2\right)} & = & \sum_{I,J,K}\sqrt{Z_{2}\left|\mathbf{i}\hat{q}_{IJK}\right|},\label{eq:l}\\
\hat{v}_{AL} & = & \sqrt{Z\left|\sum_{I<J<K}\epsilon\left(e_{I},e_{J},e_{K}\right)\mathbf{i}\hat{q}_{IJK}\right|}.\label{eq:m}
\end{eqnarray}
In this subsection, we will study the properties of the three-hand
operator $\hat{q}_{IJK}$ defined in (\ref{eq:hey}); this will be
followed by the subleties neglected in the literatures of volume operators.
Firstly, from the definition in (\ref{eq:hey}), $\hat{q}_{IJK}$
could classified into three cases, where: all the three indices are
equal, two of them are equal, and none of them are equal. The first
and second case, namely $\hat{q}_{III}$ and $\hat{q}_{IIJ},$ are
interestingly, not zero. By a direct calculation from (\ref{eq:hey}),
one could obtain:
\begin{eqnarray*}
\hat{q}_{III} & = & 4\left|J_{I}\right|^{2},\\
\hat{q}_{IIJ} & = & 4J_{I}\cdot J_{J},
\end{eqnarray*}
where both operators are clearly hermitian (real, symmetric) since
the angular momentum $J_{I}$ is hermitian, with respect to the inner
product defined on the Hilbert space. The second case, particularly
has the following symmetries: 
\begin{eqnarray*}
\hat{q}_{IIJ} & =- & \hat{q}_{IJI}=\hat{q}_{JII}=\hat{q}_{IJJ}.
\end{eqnarray*}
The last case is the condition where the indices $I,J,K$ are distinct.
With this condition, (\ref{eq:hey}) could be written as (\ref{eq:Vrs}),
which is an antihermitian (real, antisymmetric) matrix, since the
commutator of distinct components of angular momentum is antihermitian.
It has the following symmetries: 
\begin{eqnarray*}
\hat{q}_{IJK} & = & \hat{q}_{JKI}=\hat{q}_{KIJ}=-\hat{q}_{JIK}=-\hat{q}_{KJI}=-\hat{q}_{IKJ}.
\end{eqnarray*}
We found that the hermiticity of three-hand operator (\ref{eq:hey})
depends on the indices $I,J,K$. Using the symmetry properties of
these three case, one could prove the following relation: 
\begin{equation}
\sum_{I,J,K=1}^{N}\hat{q}_{IJK}=\sum_{I=1}^{N}4\left|J_{I}\right|^{2}.\label{eq:nol}
\end{equation}

It should be kept in mind that since $\hat{q}_{III}$ and $\hat{q}_{IIJ}$
are symmetric, the imaginary factor $\mathbf{i}$ on the volume operator
alter them to their antihermitian counterparts, while for antisymmetric
$\hat{q}_{IJK}$ as shown in \cite{Carlo4,Thimanclosed,Brunemann1},
it becomes hermitian. Thanks to the existence of the absolute value
in the operator, all the $\mathbf{i}\hat{q}_{III}$ and $\mathbf{i}\hat{q}_{IIJ}$
terms becomes hermitian, which is the consequence from Claim 4. Moreover
they satisfy: 
\begin{eqnarray*}
\left|\mathbf{i}\hat{q}_{IIJ}\right|=\left|\mathbf{i}\hat{q}_{IJJ}\right|, & \qquad & \left|\mathbf{i}\hat{q}_{IJK}\right|=\left|\mathbf{i}\hat{q}_{IKJ}\right|.
\end{eqnarray*}
Using this results, we could show that the Rovelli-Smolin volume operator
(\ref{eq:rs1}) and (\ref{eq:rs2}), could we written as: 
\begin{equation}
\hat{v}_{RS}^{\left(1\right)}=\sqrt{Z_{1}\left(\sum_{I=1}^{N}\left|\mathbf{i}\hat{q}_{III}\right|+3!\sum_{I<J}^{N}\left|\mathbf{i}\hat{q}_{IIJ}\right|+3!\sum_{K<L<M}^{N}\left|\mathbf{i}\hat{q}_{KLM}\right|\right)},\label{eq:vrs1a}
\end{equation}
\begin{equation}
\hat{v}_{RS}^{\left(2\right)}=\sum_{I=1}^{N}\sqrt{Z_{2}\left|\mathbf{i}\hat{q}_{III}\right|}+3!\sum_{I<J}^{N}\sqrt{Z_{2}\left|\mathbf{i}\hat{q}_{IIJ}\right|}+3!\sum_{K<L<M}^{N}\sqrt{Z_{2}\left|\mathbf{i}\hat{q}_{KLM}\right|}.\label{eq:vr2a}
\end{equation}
However, following the argument derived in \cite{Carlo4}, as the
three-hand operator acts on the nodes of a spin-network $\Gamma$
\cite{Eugenio,Carlo4}, the linearly dependent terms $\mathbf{i}\hat{q}_{III}$
and $\mathbf{i}\hat{q}_{IIJ}$ gives zero contributions to the sum,
since, in the graphical representation of spin-network, only terms
in which each hand of the operator 'grasps' a distinct links give
non-vanishing contributions. Another argument which strengthen the
removal of the linearly dependent terms by hand is the fact that the
volume operator acting on gauge-invariant trivalent graph must be
zero since it is a planar graph \cite{Renata}. It is clear that in
the angular-momentum representation of spin-network, the action of
(\ref{eq:vrs1a}) and (\ref{eq:vr2a}) on a planar graph is not zero
by the existence of the linearly dependent terms, thus giving different
results with the graphical representation. Therefore, by consciously
removing the linearly dependent terms due to the reason explained
previously, the Rovelli-Smolin operator are exactly written as (\ref{eq:rs1})
and (\ref{eq:rs2}). Nevertheless, the ambiguities on the regularization
procedure of the Rovelli-Smolin operator is dicussed in detailed in
\cite{kristina1,Kristina2}.

As for the Ashtekar-Lewandowski volume operator, the existence of
the sign term $\epsilon\left(e_{I},e_{J},e_{K}\right)$ is crucial
for three reasons: (1) Without the existence of $\epsilon\left(e_{I},e_{J},e_{K}\right)$,
the Ashtekar-Lewandowski volume operator (\ref{eq:huf}) will also
encounter the similar problem with the Rovelli-Smolin operator, namely,
the non-vanishing of linearly dependent terms, due to condition (\ref{eq:nol}).
(2) The existence of $\epsilon\left(e_{I},e_{J},e_{K}\right)$ removes
the linearly dependent terms from the summation since parallel $e_{I}$'s
will give zero, and this cause the possibility to write (\ref{eq:huf})
as (\ref{eq:AL}). (3) Its existence also guarantees that $\hat{v}_{AL}$
gives the same result either in the angular momentum or the graphical
representation of spin-network.

\subsection{Arguments for the Hermiticity of the Volume Operators}

From the previous section, we have found that the hermiticity of $\hat{q}_{IJK}$
depends on the indices $I,J,K$. For distinct $I,J,K,$ $\hat{q}_{IJK}$
is antisymmetric, therefore, it is clear that the arguments inside
the absolute value in the volume operators (\ref{eq:k}), (\ref{eq:l})
and (\ref{eq:m}) are hermitian (for the case of $\hat{v}_{AL}$,
the quantity inside the absolute value is only a linear combination
of hermitian matrices, which is still hermitian). Let us call these
hermitian quantities as $M\sim\mathbf{i}q$. Regardless of the hermiticity
of $M$, the quantity $\left|M\right|$ is guaranteed to be hermitian
and positive semi-definite, by Claim 4. Thus one could conclude further
that the arguments inside the square root in (\ref{eq:k}), (\ref{eq:l})
and (\ref{eq:m}) are hermitian and positive semi-definite (for the
case of $\hat{v}_{RS}^{\left(1\right)}$, the hermiticity and non-negativity
are closed under addition) since the constant $Z$'s are positive.
Finally, using Claim 3, it is clear that the volume operators (\ref{eq:k}),
(\ref{eq:l}) and (\ref{eq:m}) are hermitian, and positive semi-definite
(where the closure of hermiticity and positive semi-definiteness under
addition is again used for the case of $\hat{v}_{RS}^{\left(2\right)}$).
This ends the proof of the hermiticity of the volume operators in
LQG. 

It needs to be kept in mind that the hermiticity (and the non-negativity
as well) of the operators is defined with respect to the inner product
defined on the Hilbert space of the graph, namely $\mathcal{H}_{\Gamma}=L_{2}\left[\textrm{SU(2)},\textrm{d}\mu_{\textrm{Haar}}\right],$
which could be extended to the Ashtekar-Lewandowski Hilbert space,
$\mathcal{H}_{AL}=L_{2}\left[\mathcal{A},\textrm{d}\mu_{\textrm{AL}}\right],$
$\mathcal{A}$ is the functional space of $\mathfrak{su(2)}$-valued
1-form and $\textrm{d}\mu_{\textrm{AL}}$ is the Ashtekar-Lewandowski
measure \cite{Ashtekar4,Ashtekar5}.

Furthermore, from definition (\ref{eq:abs}), one could show that:
\begin{equation}
\mathbf{x}^{\dagger}\left|\mathbf{i}\hat{q}_{IJK}\right|\mathbf{x}=\mathbf{x}^{\dagger}\sqrt{-\hat{q}_{IJK}^{2}}\mathbf{x}=\mathbf{x}^{\dagger}\sqrt{-\hat{q}_{IJK}\hat{q}_{IJK}}\mathbf{x}=\mathbf{x}^{\dagger}\sqrt{\hat{q}_{IJK}^{T}\hat{q}_{IJK}}\mathbf{x}\geq0,\label{eq:mutlak}
\end{equation}
where the last equality comes from the antisymmetricity of $\hat{q}_{IJK}.$
Therefore the volume operators (\ref{eq:k}), (\ref{eq:l}) and (\ref{eq:m})
could be written as: 
\begin{eqnarray}
\hat{v}_{RS}^{\left(1\right)} & = & Z_{1}^{\nicefrac{1}{2}}\sqrt{\overline{\sum_{I,J,K}}\sqrt{\hat{q}_{IJK}^{T}\hat{q}_{IJK}}},\label{eq:k1}\\
\hat{v}_{RS}^{\left(2\right)} & = & \overline{\sum_{I,J,K}}Z_{2}^{\nicefrac{1}{2}}\sqrt[4]{\hat{q}_{IJK}^{T}\hat{q}_{IJK}},\label{eq:l1}\\
\hat{v}_{AL} & = & Z^{\nicefrac{1}{2}}\sqrt[4]{-\left(\sum_{I<J<K}\epsilon\left(e_{I},e_{J},e_{K}\right)\hat{q}_{IJK}\right)^{2}}.\label{eq:m11}
\end{eqnarray}
Since for distinct $I,J,K,$ $\hat{q}_{IJK}$ is real and antisymmetric
as proven in \cite{Carlo4,Thimanclosed,Brunemann1}, $\hat{q}_{IJK}^{T}\hat{q}_{IJK}$
is therefore real, symmetric and positive semi-definite (in fact,
this statement could be proven using a similar procedure used to prove
Claim 1, by replacing $\mathbb{C}^{n}\rightarrow\mathbb{R}^{n}$ and
$\dagger\rightarrow T$). Moreover, one could convince her/himself
that $\hat{q}_{IJK}^{T}\hat{q}_{IJK}$ admits a unique real, symmetric,
and positive semi-definite square root (in fact, this statement is
the real version of Claim 4, which could be proven by a similar manner),
namely $\sqrt{\hat{q}_{IJK}^{T}\hat{q}_{IJK}}$, equivalent to $\left|\mathbf{i}\hat{q}_{IJK}\right|$
by relation (\ref{eq:mutlak}). This prove that $\left|\mathbf{i}\hat{q}_{IJK}\right|$
is real, symmetric and positive semi-definite. Finally, realizing
that symmetricity and non-negativity are closed under addition, we
could conclude that, in the angular-momentum representation of spin-network,
the Rovelli-Smolin and the Ashtekar-Lewandowski volume operator are
real, symmetric, and positive semi-definite with respect to the inner
product defined on $\mathcal{H}_{\Gamma}=L_{2}\left[\textrm{SU(2)},\textrm{d}\mu_{\textrm{Haar}}\right]$. 

As mentioned in the Introduction, the symmetricity of the Rovelli-Smolin
volume operator had been proven in \cite{Carlo4}, based on the graphical
representation of the spin-network. In this approach \cite{Carlo4,JinsongYang},
one could obtain the matrix representation of an operator without
fixing a notion of inner product. Therefore, the reality and symmetricity
of the area and volume operator obtained from the graphical approach,
could be used to fix an inner product on the representation space
of the graph, such that it matches the inner product induced by the
Haar measure on SU(2) \cite{Carlo4}. Our work in this article could
be interpreted as a complement to the result obtained in \cite{Carlo4};
in which given a Hilbert space $\mathcal{H}_{\Gamma}=L_{2}\left[\textrm{SU(2)},\textrm{d}\mu_{\textrm{Haar}}\right]$,
we prove, based on the angular-momentum approach, that volume operators
acting on $\mathcal{H}_{\Gamma}$ are real, symmetric, and positive
semi-definite, with respect to the inner product induced by the Haar
measure. The result gives a positive argument on the equivalency between
the graphical and the angular momentum approach of spin-network. 

The reality, symmetricity, and positive semi-definiteness of an operator
are already some advantages, hence for the volume operators case,
we could reveal another interesting property. One notice that the
symmetric matrix comes from a product of a purely imaginary antisymmetric
matrix with its complex-conjugate as in (\ref{eq:mutlak}). The eigenvalues
of this type of matrix always come in pairs, namely $\pm\lambda$
for even dimension, with an additional zero for the odd dimension.
Therefore the resulting symmetric matrix always have positive degenerate
eigenvalues coming in pairs $\lambda_{1,2}^{2},$ with an additional
zero if the dimension is odd. One could prove that for a real symmetric
matrix, the eigenspaces of distinct eigenvalues are orthogonal to
each other. For repeated eigenvalues with mutiplicity $m$, one could
obtain $m$ orthonormal eigenvectors which spans the eigensubspace.
These orthonormal eigenvectors can be choosen arbitrarily in the sense
that they are invariant up to an orthogonal transformation in $m$-dimension.
Moreover, the eigenvectors of real, symmetric matrices can always
be choosen to be real. We would like to stress that there exists a
freedom in choosing the orthonormal eigenvectors for each 2-dimensional
eigensubspaces as a consequence of the degeneracies in the eigenvalues.
The eigenvectors do not necessarily need to be the ones which correspond
to the non-degenerate eigenvalues pair $\pm\lambda$ of $\mathbf{i}\hat{q}_{IJK}$,
as suggested in the following literatures \cite{Eugenio,Thimanclosed,Brunemann1,Brunemann2}.

\section{The Matrix Representation of the Volume Operator}

On a series of remarkable articles, Thiemann and Brunemann gave an
explicit closed formula for the matrix $\hat{q}_{IJK},$ which are
given as follows. Let $\left|\vec{a}\left(12\right)\right\rangle $
be a state of $N$-coupled spins, written in the recoupling scheme
basis \cite{Brunemann1}:
\[
\left|\vec{g}\left(IJ\right)\right\rangle =\left|\vec{g}\left(IJ\right),\vec{j},j=0,m=0\right\rangle ,
\]
which diagonalize the following complete sets of commuting observables:
\[
\left|J\right|^{2}=\left|\sum_{I=1}^{N}J_{I}\right|^{2},\quad J^{3},\quad\left|J_{I}\right|^{2},I=1,..,N,
\]
and the modulus of the following $N-2$ operators (let us suppose
for the moment $I,J\neq1,2$), with their corresponding spins eigenvalues
are:
\[
\begin{array}{ccccccc}
G_{1} & := & J_{1},\quad\quad\quad\; & \qquad & g_{1} & = & j_{1},\quad\quad\quad\;\\
G_{2} & := & G_{1}+J_{J},\quad &  & g_{2} & = & g_{2}\left(g_{1},j_{J}\right),\\
G_{3} & := & G_{2}+J_{1},\quad &  & g_{3} & = & g_{3}\left(g_{2},j_{1}\right),\\
G_{4} & := & G_{3}+J_{2},\quad &  & g_{4} & = & g_{4}\left(g_{3},j_{2}\right),\\
 & \cdots &  &  &  & \cdots\\
G_{I} & := & G_{I-1}+J_{I-2}, &  & g_{I} & = & g_{I}\left(g_{I-1},j_{I-2}\right),\quad\\
G_{I+1} & := & G_{I}+J_{I-1},\quad &  & g_{I+1} & = & g_{I+1}\left(g_{I},j_{I-1}\right),\quad\\
G_{I+2} & := & G_{I+1}+J_{I+1}, &  & g_{I+2} & = & g_{I+2}\left(g_{I+1},j_{I+1}\right),\\
G_{I+3} & := & G_{I+2}+J_{I+2}, &  & g_{I+3} & = & g_{I+3}\left(g_{I+2},j_{I+2}\right),\\
 & \cdots &  &  & \cdots\\
G_{J} & := & G_{J-1}+J_{J-1}, &  & g_{J} & = & g_{J}\left(g_{J-1},j_{J-1}\right).\quad
\end{array}
\]
The matrix representation of $\hat{q}_{IJK}$ in this basis is: 
\begin{eqnarray}
\left\langle \vec{a}\left(12\right)\right|\hat{q}_{IJK}\left|\vec{a}'\left(12\right)\right\rangle  & = & \frac{1}{4}\left(-1\right)^{j_{I}+j_{K}+a_{I-1}+a_{K}}\left(-1\right)^{a_{I}-a'_{I}}\left(-1\right)^{\sum_{n=I+1}^{J-1}j_{n}}\left(-1\right)^{-\sum_{p=J+1}^{K-1}j_{p}}\prod_{n=2}^{I-1}\delta_{a_{n}a'_{n}}\prod_{n=K}^{N}\delta_{a_{n}a'_{n}}\times\nonumber \\
 &  & \times\; X\left(j_{I},j_{J}\right)^{\nicefrac{1}{2}}X\left(j_{J},j_{K}\right)^{\nicefrac{1}{2}}\sqrt{\left(2a_{I}+1\right)\left(2a'_{I}+1\right)}\sqrt{\left(2a_{J}+1\right)\left(2a'_{J}+1\right)}\times\nonumber \\
 &  & \times\;\left\{ \begin{array}{ccc}
a_{I-1} & j_{I} & a_{I}\\
1 & a'_{I} & j_{I}
\end{array}\right\} \left(\prod_{n=I+1}^{J-1}\sqrt{\left(2a_{n}+1\right)\left(2a'_{n}+1\right)}\left(-1\right)^{a_{n-1}+a'_{n-1}+1}\left\{ \begin{array}{ccc}
j_{n} & a'_{n-1} & a'_{n}\\
1 & a{}_{n} & a_{n-1}
\end{array}\right\} \right)\times\nonumber \\
 &  & \times\;\left\{ \begin{array}{ccc}
a_{K} & j_{K} & a_{K-1}\\
1 & a'_{K-1} & j_{K}
\end{array}\right\} \left(\prod_{n=J+1}^{K-1}\sqrt{\left(2a_{n}+1\right)\left(2a'_{n}+1\right)}\left(-1\right)^{a_{n-1}+a'_{n-1}+1}\left\{ \begin{array}{ccc}
j_{n} & a'_{n-1} & a'_{n}\\
1 & a{}_{n} & a_{n-1}
\end{array}\right\} \right)\times\nonumber \\
 &  & \times\;\left(\left(-1\right)^{a'_{J}+a'_{J-1}}\left\{ \begin{array}{ccc}
a_{J} & j_{J} & a'_{J-1}\\
1 & a{}_{J-1} & j_{J}
\end{array}\right\} \left\{ \begin{array}{ccc}
a'_{J-1} & j_{J} & a'_{J}\\
1 & a{}_{J} & j_{J}
\end{array}\right\} \right.-..\nonumber \\
 &  & \qquad\qquad\left...-\left(-1\right)^{a{}_{J}+a{}_{J-1}}\left\{ \begin{array}{ccc}
a'_{J} & j_{J} & a'_{J-1}\\
1 & a{}_{J-1} & j_{J}
\end{array}\right\} \left\{ \begin{array}{ccc}
a{}_{J-1} & j_{J} & a'_{J}\\
1 & a{}_{J} & j_{J}
\end{array}\right\} \right),\label{eq:qijk}
\end{eqnarray}
with $X\left(j_{I},j_{J}\right)=2j_{I}\left(2j_{I}+1\right)\left(2j_{I}+2\right)2j_{J}\left(2j_{J}+1\right)\left(2j_{J}+2\right)$
and $\left|\vec{a}\left(12\right)\right\rangle =\left|\vec{a}\left(12\right),\vec{j},j=0,m=0\right\rangle $.
In this step, one use a specific inner product $\left\langle ..,..\right\rangle $
to obtain the matrix components. Using this formula, one can obtain
the matrix representation of the volume operator, which is need to
be done with an extra care. Nevertheless, it needs to be kept in mind
that it is not possible to obtain a closed analytical formula for
the matrix representation of the volume operator, due to the complicated
nature of (\ref{eq:qijk}).

The gauge invariance is implemented by setting the total angular momentum
operator to be zero \cite{Brunemann1}: 
\begin{equation}
J\left|\psi\right\rangle =\sum_{i=1}^{N}J_{i}\left|\psi\right\rangle =0.\label{eq:gauss}
\end{equation}
Projecting the operator to the kinematical Hilbert space $\mathcal{K}_{\Gamma}\subset\mathcal{H}_{\Gamma}$
defined by (\ref{eq:gauss}), they satisfy the following condition:
\begin{equation}
J=\sum_{i=1}^{N}J_{i}=G_{N}=G_{N-1}+J_{N}\approx0.\label{eq:gauge}
\end{equation}
This cause a recursive constraints to occur on each steps of the spins
addition: 
\begin{eqnarray}
G_{N-1} & = & G_{N-2}+J_{N-1}\approx-J_{N}\label{eq:g}\\
G_{N-2} & = & G_{N-3}+J_{N-2}\approx-J_{N}-J_{N-1}.\nonumber 
\end{eqnarray}
The equivalence notation $\approx$ is used to denoted that these
relations are valid only on the constrained Hilbert space. Therefore,
the spin quantum number of spin-network state are also restricted:
\begin{equation}
\textrm{max}\left(\left|j_{N-2}-g_{N-3}\right|,\left|j_{N}-j_{N-1}\right|\right)\leq g_{N-2}\leq\textrm{min}\left(\left|j_{N-2}+g_{N-3}\right|,\left|j_{N}+j_{N-1}\right|\right).\label{eq:batas}
\end{equation}
An important thing we need to stress is that relation (\ref{eq:gauge})-(\ref{eq:g})
are only valid on the kinematical Hilbert space $\mathcal{K}_{\Gamma}$.
Acting on the full, non-gauge invariant Hilbert space $\mathcal{H}_{\Gamma}$,
the quantum gauge invariance is not restricting the spin operators
to satisfy (\ref{eq:gauge})-(\ref{eq:g}), but to select the set
of invariant states such that they satisfies the Gauss constraint,
and thus defines $\mathcal{K}_{\Gamma}$.

The formal steps to obtain the matrix representation of volume operators
in the angular momentum representation of the spin-network could be
summarize as follows: 
\begin{enumerate}
\item Starting from (\ref{eq:k1}), (\ref{eq:l1}), and/or (\ref{eq:m11}),
one finds the $\hat{q}_{IJK}^{T}\hat{q}_{IJK}$ for $\hat{v}_{RS}$
and/or the terms under the 4-root for $\hat{v}_{AL}$. 
\item The next step is to diagonalize the terms under the root to obtain
the spectrum, particularly the pair $\left(\Lambda,\lambda_{n}P_{n}\right),$
with $\lambda_{n}P_{n}$ is the diagonal matrix of $\hat{q}_{IJK}^{T}\hat{q}_{IJK}$,
and $\Lambda$ is the corresponding orthogonal transformation. In
this step one should check that all the eigenvalues are positive. 
\item Following similar procedure as in Claim 3, one takes the 4-root of
$\lambda_{n}P_{n}$ for $\hat{v}_{RS}^{\left(2\right)}$ and $\hat{v}_{AL}$
(or the square root for $\hat{v}_{RS}^{\left(1\right)}$) by simply
4-rooting (or square-rooting for $\hat{v}_{RS}^{\left(1\right)}$)
the eigenvalues $\lambda_{n}$ to obtain $+\sqrt[4]{\lambda_{n}P_{n}}$
(or +$\sqrt{\lambda_{n}P_{n}}$ for $\hat{v}_{RS}^{\left(1\right)}$). 
\item The next step is to de-diagonalize the matrix resulting from Step
3: $\Lambda^{-1}\sqrt[4]{\lambda_{n}P_{n}}\Lambda$ to obtain $\sqrt[4]{\hat{q}_{IJK}^{T}\hat{q}_{IJK}}$
for $\hat{v}_{RS}^{\left(2\right)}$ (and the same procedure for $\hat{v}_{AL}$),
while for $\hat{v}_{RS}^{\left(1\right)}$, one needs to de-diagonalize
the following matrix $\Lambda^{-1}\sqrt{\lambda_{n}P_{n}}\Lambda$
to obtain $\sqrt{\hat{q}_{IJK}^{T}\hat{q}_{IJK}}$, then to sum over
distinct $I,J,K,$ and finally to repeat once again Step 4.
\item In this step, one already obtain the matrix representation of volume
operators (\ref{eq:k1}), (\ref{eq:l1}),and/or (\ref{eq:m11}); they
are ready to be diagonalized to obtain the spectrum. 
\end{enumerate}

\subsection{The 4-Vertex Case}

The volume operator acts on the nodes of a spin-network $\Gamma$,
as already discussed in \cite{Eugenio,Carlo4}. Let us consider a
4-vertex node, where the direction of the four links labelled by $j_{I}$,
$I=1,2,3,4$, is chosen to be pointing outward from the node. The
possible three-hand operator $\hat{q}_{IJK}$ for the 4-vertex volume
operators are the ones with distinct indices, where $I,J,K$ runs
from 1 to 4. As already shown in \cite{Brunemann1}, restricting to
the kinematical Hilbert space, the possible three-hands operator for
4-vertex case satisfies the following relations: 
\[
\hat{q}_{123}\approx-\hat{q}_{124}\approx\hat{q}_{134}\approx-\hat{q}_{234}.
\]
Inserting this condition to (\ref{eq:k1}), (\ref{eq:l1}),and (\ref{eq:m11})
gives:
\begin{eqnarray}
\hat{v}_{RS}^{\left(1\right)} & = & 2Z_{1}^{\nicefrac{1}{2}}\sqrt{\left|\mathbf{i}\hat{q}_{123}\right|},\label{eq:k1-1}\\
\hat{v}_{RS}^{\left(2\right)} & = & 4Z_{2}^{\nicefrac{1}{2}}\sqrt{\left|\mathbf{i}\hat{q}_{123}\right|}=2\sqrt{\frac{Z_{2}}{Z_{1}}}\hat{v}_{RS}^{\left(1\right)},\label{eq:l1-1}\\
\hat{v}_{AL} & = & 2Z^{\nicefrac{1}{2}}\sqrt{\left|\mathbf{i}\hat{q}_{123}\right|}=\sqrt{\frac{Z_{3}}{Z_{1}}}\hat{v}_{RS}^{\left(1\right)}.\label{eq:m11-1}
\end{eqnarray}
The three volume operators give the same expression and only differ
by the constants.

For a 4-vertex case, (\ref{eq:qijk}) greatly simplifies into:
\begin{eqnarray}
\left\langle j_{12}\right|\hat{q}_{123}\left|j_{12}-1\right\rangle  & = & \frac{1}{\left(\left(2j_{12}-1\right)\left(2j_{12}+1\right)\right)^{\nicefrac{1}{2}}}\left(\left(j_{1}+j_{2}+j_{12}+1\right)\left(-j_{1}+j_{2}+j_{12}\right)\left(j_{1}-j_{2}+j_{12}\right)\left(j_{1}+j_{2}-j_{12}+1\right)\times..\right.\nonumber \\
 &  & \qquad\qquad\qquad..\times\left.\left(j_{3}+j_{4}+j_{12}+1\right)\left(-j_{3}+j_{4}+j_{12}\right)\left(j_{3}-j_{4}+j_{12}\right)\left(j_{3}+j_{4}-j_{12}+1\right)\right)^{\nicefrac{1}{2}},\label{eq:4vert}
\end{eqnarray}
where we write $\left|\vec{a}\left(12\right)\right\rangle =\left|a_{2}=j_{12}\right\rangle ,$
or simply $\left|j_{12}\right\rangle $. Moreover, $\left|a'_{2}=j'_{12}\right\rangle =\left|j_{12}-1\right\rangle ,$
where the detailed explanation is discussed in \cite{Brunemann1}.
The possible value of $j_{12}$ needs to satisfy the following restrictions
coming from (\ref{eq:batas}):
\[
\textrm{max}\left(\left|j_{1}-j_{2}\right|,\left|j_{3}-j_{4}\right|\right)\leq j_{12}\leq\textrm{min}\left(\left|j_{1}+j_{2}\right|,\left|j_{3}+j_{4}\right|\right).
\]
Therefore, the dimension $k$ of the intertwinner space (or kinematical
space $\mathcal{K}_{\Gamma}$), which is the dimension of the matrix
representation of $\left\langle j_{12}\right|\hat{q}_{123}\left|j_{12}-1\right\rangle $
is $k=j_{12}^{\textrm{max}}-j_{12}^{\textrm{min}}+1$. Sorting the
possible value of $j_{12}$ from the smallest one, namely $j_{12}^{\textrm{min}}$,
and using the following notation: $a_{k}=\left\langle j_{12}^{\textrm{min}}+k\right|\hat{q}_{123}\left|j_{12}^{\textrm{min}}+k-1\right\rangle $,
one could construct the matrix representation of $\hat{q}_{123}$,
which is a type of Jacobi matrix with banded structure \cite{Brunemann1}:
\[
\hat{q}_{123}=\left[\begin{array}{ccccc}
0 & -a_{1} & 0 & \cdots & 0\\
a_{1} & 0 & -a_{2} &  & \vdots\\
0 & a_{2} & 0 &  & \vdots\\
\vdots &  &  & \ddots & -a_{n-1}\\
0 & \cdots & \cdots & a_{n-1} & 0
\end{array}\right].
\]
We found that for 4-vertex case, the product of $\hat{q}_{123}$ with
its transpose could always be written as the following matrix: 
\[
\hat{q}_{123}^{T}\hat{q}_{123}=\left[\begin{array}{ccccccccc}
a_{1}^{2} & 0 & -a_{1}a_{2} & 0 & 0 & 0 & \cdots & \cdots & 0\\
0 & a_{1}^{2}+a_{2}^{2} & 0 & -a_{2}a_{3} & 0 & 0 & \cdots &  & \vdots\\
-a_{1}a_{2} & 0 & a_{2}^{2}+a_{3}^{2} & 0 & -a_{3}a_{4} & 0\\
0 & -a_{2}a_{3} & 0 & a_{3}^{2}+a_{4}^{2} & 0 & -a_{4}a_{5} & 0 & \cdots\\
0 & 0 & -a_{3}a_{4} & 0 & \ddots & 0 & \vdots &  & \vdots\\
\vdots & \vdots & 0 & -a_{4}a_{5} &  & \vdots & 0 & -a_{n-3}a_{n-2} & 0\\
\vdots &  & \vdots & 0 & \vdots & 0 & a_{n-3}^{2}+a_{n-2}^{2} & 0 & -a_{n-2}a_{n-1}\\
\vdots &  &  & \vdots & 0 & -a_{n-3}a_{n-2} & 0 & a_{n-2}^{2}+a_{n-1}^{2} & 0\\
0 & \ldots & \ldots &  & 0 & 0 & -a_{n-2}a_{n-1} & 0 & a_{n-1}^{2}
\end{array}\right],
\]
which is real, symmetric, and positive semi-definite, with respect
to the inner product defined in (\ref{eq:qijk}). With these simplification,
one could proceed to the formal steps to obtain the matrix representation
of volume operators and their spectrum.

\subsection{Monochromatic  Quantum Tetrahedron}

A 4-vertex node is geometrically interpreted as a quantum tetrahedron.
Here we give, as examples, the explicit calculation of the matrix
representation of volume operators and their spectrum for 4-vertex
in two cases: (1) Ground State Monochromatic (GSM) case, where the
spin $j_{1}=j_{2}=j_{3}=j_{4}=\frac{1}{2}$, and the First Excited
Monochromatic (FEM) case with $j_{1}=j_{2}=j_{3}=j_{4}=1$. Since
for 4-vertex case the three volume operators only differ by the constants,
we will only calculate the $\hat{v}_{RS}^{\left(1\right)}$ for brevity.

\subsubsection{Ground State Monochromatic (GSM) Case}

Using (\ref{eq:mutlak}), (\ref{eq:k1-1}) becomes: 
\[
\hat{v}_{RS}^{\left(1\right)}=2Z_{1}^{\nicefrac{1}{2}}\sqrt[4]{\hat{q}_{123}^{T}\hat{q}_{123}},
\]
and from (\ref{eq:4vert}), we obtain, respectively: 
\[
\hat{q}_{123}=\left[\begin{array}{cc}
0 & \sqrt{3}\\
-\sqrt{3} & 0
\end{array}\right],\qquad\hat{q}_{123}^{T}\hat{q}_{123}=\left[\begin{array}{cc}
3 & 0\\
0 & 3
\end{array}\right],\qquad\left|\mathbf{i}\hat{q}_{123}\right|=\left[\begin{array}{cc}
\sqrt{3} & 0\\
0 & \sqrt{3}
\end{array}\right].
\]
Since the matrix is already diagonal (in 2-dimension, the degrees
of freedom for an antisymmetric matrix is one, therefore, $\hat{q}_{123}^{T}\hat{q}_{123}$
is already diagonal in this step. This, in general, is not the case
for larger dimensional matrix), then we can directly obtain the unique
positive square root: 
\[
\hat{v}_{RS}^{\left(1\right)}=2Z_{1}^{\nicefrac{1}{2}}\left[\begin{array}{cc}
\sqrt[4]{3} & 0\\
0 & \sqrt[4]{3}
\end{array}\right],
\]
so that the eigenvalues are $\lambda_{1,2}=2Z_{1}^{\nicefrac{1}{2}}\sqrt[4]{3},$
giving exactly the same result with the procedure describe in \cite{Eugenio,Brunemann1,Brunemann2}.
The eigenvalues are degenerate, therefore one could freely choose
two orthonormal eigenvectors by the Gramm-Schmidt procedure for the
volume operator: 
\begin{eqnarray}
\left|\mathbf{n}_{1}\right\rangle =\cos\theta\left|a_{12}\left(0\right)\right\rangle -\sin\theta\left|a_{12}\left(1\right)\right\rangle , & \quad & \left|\mathbf{n}_{2}\right\rangle =\sin\theta\left|a_{12}\left(0\right)\right\rangle +\cos\theta\left|a_{12}\left(1\right)\right\rangle .
\end{eqnarray}
The freedom is described by a parameter $\theta$, which correspond
to the angle of rotation in the $2$-dimensional eigenspace.

\subsubsection{First Excited Monochromatic (FEM) Case}

From (\ref{eq:4vert}) we obtain, respectively: 
\[
\hat{q}_{123}=\left(\begin{array}{ccc}
0 & -8\frac{\sqrt{3}}{3} & 0\\
8\frac{\sqrt{3}}{3} & 0 & -4\frac{\sqrt{5}}{\sqrt{3}}\\
0 & 4\frac{\sqrt{5}}{\sqrt{3}} & 0
\end{array}\right),\qquad\hat{q}_{123}^{T}\hat{q}_{123}=\left(\begin{array}{ccc}
\frac{64}{3} & 0 & -\frac{32\sqrt{5}}{3}\\
0 & 48 & 0\\
-\frac{32\sqrt{5}}{3} & 0 & \frac{80}{3}
\end{array}\right).
\]
In contrast with the GSM case, $\hat{q}_{123}^{T}\hat{q}_{123}$ is
not diagonal; one needs to perform the following diagonalization,
to obtain the root: 
\[
\Lambda^{-1}\hat{q}_{123}^{T}\hat{q}_{123}\Lambda=\lambda_{n}\hat{P}_{n}=\left(\begin{array}{ccc}
48 & 0 & 0\\
0 & 48 & 0\\
0 & 0 & 0
\end{array}\right),\qquad\sqrt[4]{\lambda_{n}\hat{P}_{n}}=\left(\begin{array}{ccc}
2\sqrt[4]{3} & 0 & 0\\
0 & 2\sqrt[4]{3} & 0\\
0 & 0 & 0
\end{array}\right).
\]
De-diagonalizing the root using $\Lambda\sqrt[4]{\lambda_{n}P_{n}}\Lambda^{-1}$,
one obtains, respectively: 
\[
\sqrt[4]{\hat{q}_{123}^{T}\hat{q}_{123}}=\left(\begin{array}{ccc}
\frac{8}{9}\sqrt[4]{3} & 0 & -\frac{4\sqrt{5}}{9}\sqrt[4]{3}\\
0 & 2\sqrt[4]{3} & 0\\
-\frac{4\sqrt{5}}{9}\sqrt[4]{3} & 0 & \frac{10}{9}\sqrt[4]{3}
\end{array}\right),\qquad\hat{v}_{RS}^{\left(1\right)}=\sqrt{Z_{1}}\left(\begin{array}{ccc}
\frac{16}{9}\sqrt[4]{3} & 0 & -\frac{8\sqrt{5}}{9}\sqrt[4]{3}\\
0 & 4\sqrt[4]{3} & 0\\
-\frac{8\sqrt{5}}{9}\sqrt[4]{3} & 0 & \frac{20}{9}\sqrt[4]{3}
\end{array}\right).
\]
The eigenvalues are $\lambda_{0}=0$ and  $\lambda_{1,2}=4Z_{1}^{\nicefrac{1}{2}}\sqrt[4]{3}.$
$\lambda_{1,2}$ are the degenerate eigenvalues, and $\lambda_{0}$
is the 'accidental zero' in \cite{Eugenio}, which occurs due to the
oddity of the intertwiner space. The eigenvectors are:
\begin{eqnarray*}
\left|\mathbf{n}_{0}\right\rangle  & = & \frac{\sqrt{5}}{3}\left|a_{12}\left(0\right)\right\rangle +\frac{2}{3}\left|a_{12}\left(2\right)\right\rangle ,\\
\left|\mathbf{n}_{1}\right\rangle  & = & \cos\theta\left(\frac{\sqrt{2}}{3}\left|a_{12}\left(0\right)\right\rangle -\frac{\sqrt{10}}{6}\left|a_{12}\left(2\right)\right\rangle \right)-\sin\theta\left|a_{12}\left(1\right)\right\rangle ,\\
\left|\mathbf{n}_{2}\right\rangle  & = & \sin\theta\left(\frac{\sqrt{2}}{3}\left|a_{12}\left(0\right)\right\rangle -\frac{\sqrt{10}}{6}\left|a_{12}\left(2\right)\right\rangle \right)+\cos\theta\left|a_{12}\left(1\right)\right\rangle .
\end{eqnarray*}

\section{Conclusions.}

Different regularization scheme applied to the volume of a 3D region
gives two types of volume operators in LQG: the Rovelli-Smollin ($\hat{v}_{RS}$)
and Ashtekar-Lewandowski ($\hat{v}_{AL}$) volume operator. In the
literatures, there exists two types of $\hat{v}_{RS}$ as had been
discussed in Section II, which differ only by the location of the
summation with respect to the square root. Based on the angular momentum
representation of the spin-network, a careful analysis on the three-hand
operator $\hat{q}_{IJK}$ shows that the hermiticity of this operator
depends on the indices $I,J,K$. It follows that the linearly dependent
terms $\hat{q}_{III}$ and $\hat{q}_{IIJ}$ are real and symmetric
(hence hermitian), and in general do not give zero contribution to
the volume operators. For $\hat{v}_{RS}$, these linearly dependent
term is removed by the argument stemming from the graphical representation
of spin-networks, while for $\hat{v}_{AL}$, they are removed by the
existence of the sign factor $\epsilon\left(e_{I},e_{J},e_{K}\right)$.
The later is more natural in the sense that it gives consistent result,
either in the angular momentum or graphical approach, without additional
assumptions. 

The formal and rigorous proof for the hermicity and positive semi-definiteness
of the volume operators had not been given explicitly in the literatures.
In this article, we had provided a proof for these two conditions.
The first step is to provide a clear definition on the absolute value
of an operator, or particularly, for our case, the absolute value
of a matrix. A mathematical definition of the absolute value of a
bounded operator already existed, namely $\left|M\right|:=+\sqrt{M^{\dagger}M}$.
Thus we apply this definition for matrices; this is done in Section
III. Consequently, one could derive several claims. The first claim
is the hermiticity and the positive semi-definiteness of the quantity
$M^{\dagger}M$, which is followed by the non-negativity of its eigenvalues,
as proven in the second claim. By the third claim,\textit{ }$M^{\dagger}M$
admits a unique hermitian and positive semi-definite square root,
which, by definition, is $\left|M\right|$. Moreover, since $\left|M\right|$
is guaranteed to be hermitian and positive semi-definite, it also
had a hermitian and positive semi-definite square root, namely $\sqrt{\left|M\right|}$,
which is the form of matrix we expected. Therefore, considering that
each volume operator contains the square root of the absolute value
of a matrix, we could use the claims to prove that volume operators
in LQG are hermitian and positive semi-definite. Furthermore, since
the arguments inside the absolute value of the operators, namely $M$,
are purely imaginary and antisymmetric (that is, $M=\mathbf{i}\hat{q}_{IJK}$
for $\hat{v}_{RS}$ and its linear combination for $\hat{v}_{AL}$),
the absolute value of $M$ is the square root of the product of an
imaginary antisymmetric matrix with its complex-conjugate, namely
$\left|M\right|=\sqrt{\hat{q}_{IJK}^{T}\hat{q}_{IJK}}$, with $\hat{q}_{IJK}$
is real antisymmetric. As a consequence to this, the volume operator
$\hat{v}\sim\sqrt{\left|M\right|}$ is real, symmetric and positive
semi-definite. The reality and symmetricity (and hence, the hermiticity)
of these operators are defined with respect to the inner product defined
in the Hilbert space of the graph, $\mathcal{H}_{\Gamma}=L_{2}\left[\textrm{SU(2)},\textrm{d}\mu_{\textrm{Haar}}\right]$.
As an advantage to this fact, one could choose the corresponding eigenvectors
to be real. Moreover, the diagonalization of the volume operator $\hat{v}$
always yield pairs of positive degenerate eigenvalues with an additional
zero if the dimension is odd. As a result of this degeneracy, one
has a freedom to choose eigenbasis for each eigensubspaces.

There exists two equivalent representations of spin-network, namely,
the angular momentum and the graphical representation. In the later,
one does not necessarily fix the notion of inner product, albeit it
is possible to obtain the matrix representation of an operator. In
particular, one could obtain the reality and symmetricity of the area
and volume operator solely from the rules of the graphical approach.
With these conditions in hand, one could fix an inner product on the
representation space of the graph, such that it matches the one defined
on the Hilbert space in the angular momentum approach; this had been
done in the literatures. Our work in this article could be interpreted
as a complement to this result; in which given a Hilbert space $\mathcal{H}_{\Gamma}=L_{2}\left[\textrm{SU(2)},\textrm{d}\mu_{\textrm{Haar}}\right]$,
we prove, based on the angular-momentum approach, that volume operators
acting on $\mathcal{H}_{\Gamma}$ are real, symmetric, and positive
semi-definite, with respect to the inner product induced by the Haar
measure on SU(2). The result supports the equivalency between the
graphical and the angular momentum approach of spin-network. 

In Section IV, we provide a formal procedure to obtain the matrix
representation of volume operator, together with their spectrum. In
order to compare our procedure with the earlier ones existing in the
literatures, we give explicit computational examples for the case
of monochromatic quantum tetrahedron. The spectrum is obtained directly
from the diagonalization of the volume operators matrices. The results
are consistent with existing results found in the literatures, although
the two procedures differ in which the spectrum resulting from the
standard approach is obtained from the diagonalization of $\mathbf{i}\hat{q}_{IJK}$
rather than $\hat{v}$. This consistency gives an approval to our
procedure as a formal approach in obtaining the eigenvalues of volume
operator, based on the angular momentum representation of spin-network.
Furthermore, our approach provides a clearer view on the symmetries
of the volume operator. Since in our approach the matrices of the
volume operators are obtained explicitely, it becomes clear that they
contain degeneracies, particularly, the eigensubspaces corresponding
to each distinct positive eigenvalues are two-folds. It might be interesting
to see how this fact will affect the calculation concerning the length
operator, Hamiltonian (and Master) constraint operator, and other
operators related to the volume operator. However,
as explained in the Introduction, the Ashtekar-Lewandowski (and Rovelli-Smolin)
volume operator could be considered as a 'special' case of the polyhedral
volume operator introduced by Bianchi, Dona', and Speziale. The latter
had a correct and well-defined semi-classical limit for any number
of vertices, in contrast with the former, which only admit a correct
semi-classical limit for $n=6$. Due to this reason, it is important
to study the polyhedral volume operator in detail for the further
works, as well as obtaining its complete spectral properties.

\section*{Acknowledgment}

S. A. is supported by Institut Teknologi Bandung In House Collaboration
Post Doctoral Fellowship. F. P. Z. would like to thank Kemenristek
DIKTI Indonesia for financial supports. I. H. and I. S. would like
to thank the members of Theoretical Physics Groups of Institut Teknologi
Bandung for the hospitality.

\end{document}